\documentclass[10pt,a4paper]{article}

\usepackage[utf8]{inputenc}
\usepackage{mathtools}
\usepackage{bm}
\usepackage{graphics}
\usepackage{fancyhdr}
\usepackage{enumitem}
\usepackage{cite}

\usepackage[absolute]{textpos}

\begin{document}

\begin{center}

	\huge
	\textbf{Imaging thick samples with optical tomography}
	\vspace{0.5cm}
	
	\large
	Alexandre Goy,$^1$ Morteza H. Shoreh,$^1$ JooWon Lim,$^1$ Michael Unser,$^2$ and Demetri Psaltis$^{1,*}$
	\vspace{0.4cm}
	
	\normalsize	\textit{$^1$Laboratoire d'Optique, École Polytechnique Fédérale de Lausanne (EPFL), Lausanne, Switzerland.\\ $^2$Biomedical Imaging Group, École Polytechnique Fédérale de Lausanne (EPFL), Lausanne, Switzerland. \\ $^*$\underline{demetri.psaltis@epfl.ch}}
	\vspace{0.2cm}
	
\end{center}
\vspace{0.5cm}
\normalsize

\textbf{Abstract: Optical diffraction tomography (ODT), initially described in the seminal paper of Emil Wolf [Opt. Comm., 1(4), 153-156 (1969)], has received renewed attention recently. The vast majority of ODT experiments reported to date have been 3D images of single cells or beads demonstrating excellent image quality and sectioning capability. The sample thickness in these experiments is 10 to 20 $\mu$m and the optical thickness is generally well below 2$\pi$. In this paper, we explore ODT when the sample consists of multiple layers of cells. We assess experimentally the impact of sample thickness for the different ODT reconstruction algorithms and we describe a strategy that allows us to image, for the first time, multi-cell clusters.}

\section{Introduction}

Optical tomography~\cite{wolf:1969, lauer:2001, charriere:2006, choi:2007, choi:2008, sung:2009, cotte:2013, merola:2017, lee:2013, devaney:1981, muller:2016} and related techniques are unique tools to quantitatively measure the three-dimensional refractive index distributions of weakly absorbing samples. In general, the method relies on the measurement of the optical field scattered from a sample illuminated with a set of known incident fields. The incidence angle of the illumination plane wave is usually the parameter that is varied and the scattered field is measured for each incident wave. In the initial studies~\cite{charriere:2006}, the sample was rotated and the scattered field was collected in a transmission geometry. In more recent works, the sample is placed in a conventional microscope and the angle of illumination is changed, within the range of angles allowed by the numerical aperture of the objective lenses~\cite{lauer:2001, choi:2007, choi:2008, sung:2009}. In order to reconstruct the object from the collection of the scattered fields, a propagation model has to be assumed. The simplest model relies on the assumption of straight rays along which the phase is equal to the integral of the optical path. In that case, the Radon inverse transform or filtered-back projection can be used~\cite{radon:1917}. However, in many cases, such as in biological samples, the features of interest have sizes comparable to the optical wavelength used to probe them. Therefore, the first significant improvement over the straight ray approximation comes by taking the effect of diffraction into account. This idea was introduced in the seminal paper by Emil Wolf~\cite{wolf:1969} and led to the technique that is now widely referred to as Optical Diffraction Tomography (ODT). In the paper by Wolf, the forward model was based on the first order Born approximation that assumes the scattered field to be negligible with respect to the incident field. The first order Born approximation is equivalent to the assumption of single scattering and is therefore limited to samples that are weakly scattering. Another significant improvement was brought to the technique by Anthony Devaney who suggested the use of the Rytov approximation instead of the first Born approximation~\cite{merola:2017, lee:2013}. Like the first order Born approximation, the first order Rytov approximation is also a linearization of the inverse scattering problem but it has been found to yield superior results for biological cells and has been the most commonly used technique for linear ODT. We believe the main reason for the superiority of the Rytov model over the Born model is the phase unwrapping that is explicit in the Rytov model. This unwrapped phase is used instead of the field in the inversion formula introduced by Wolf (which we refer to as the Wolf transform). However, this substitution is the source of severe defocus-like distortions in the reconstruction for thick objects in sections far from the plane in which the measurement has been taken. In fact, due to this limitation essentially all experimental demonstrations of ODT use thin samples, typically single cells. The solution to correct this distortion is to refocus the measurement to the plane of interest before unwrapping the phase and applying the Wolf transform. This technique, referred to as the hybrid approach~\cite{devaney:2012}, has been already described in the field ultrasound imaging~\cite{sponheim:1991}, but has been only recently characterized in optics~\cite{kostencka:2016}. We refer to this method as the Refocused Rytov method. A similar work was also published in optics regarding the refocusing of the scattered field when the straight rays approximation is used~\cite{choi:2008_2}. The main limitation of both the first order Born and Rytov approximation based models is that they rely on the first order approximation, which is obviously violated in thicker and denser objects. In the Born and Rytov formalism, each slice of the object scatters the light independently. This is of course not true when the field incident on a particular slice is distorted by the slices upstream or, similarly, when the field scattered for the slice is distorted by other slices downstream. 

It has been shown recently that it is possible to incorporate multiple scattering by using a nonlinear forward model~\cite{kamilov:2015, kamilov:2016}. In this earlier work~\cite{kamilov:2015, tian:2015, kamilov:2016}, we introduced Learning Tomography (LT), a beam propagation based optimization technique that makes use of machine learning concepts. We recently showed that this method is superior in solving the inverse scattering problem posed by strongly scattering objects~\cite{lim:2017}. 

In this paper, we experimentally demonstrate the use of the Learning Tomography (LT) and the refocused Rytov techniques for thick samples consisting of clusters of yeast cells. We used agarose gels to assemble cells into samples of variable optical density and thickness for the purpose of exploring the limit of tomography techniques as thickness and scattering increases. In particular, we show that the LT algorithm produces well focused images throughout the volume of the sample when the Rytov solution that is refocused on only a subset of planes (one or several) is used as the initial condition. In addition, the LT algorithm generally yields images of superior quality, displaying less background noise than the Rytov solution and sharper features.

\section{Experimental apparatus and sample preparation}

\begin{figure}[h!]
  \begin{center}
    \includegraphics{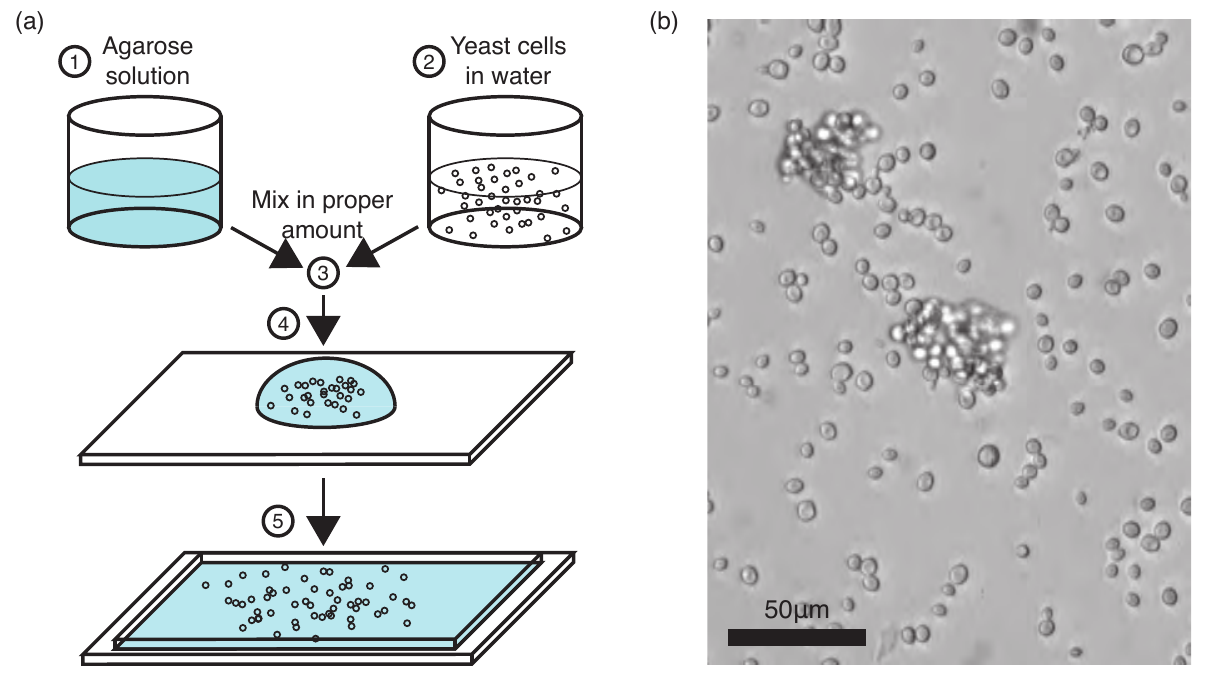}
     \caption{\small{\textbf{Sample preparation.} (a) Step 1: A solution of typically 1\% agarose is prepared and heated to 80$^\circ$C. Step 2: The yeast cells are dispersed in aqueous solution with the desired concentration. Step 3: the cells are mixed with the agarose solution when the latter has cooled down to a temperature of 35$^\circ$C. Step 4: A controlled amount of he mixed solution is put on the cover slip using a micro-pipette. The amount of liquid determines the thickness of the finished sample. Step 5: A 150 micron thick cover slip is placed on the solution that spreads in-between the slides. The two slides are then sealed using epoxy glue. (b) Wide field transmission (incoherent white light) image of a typical sample of yeast cells clusters mounted between cover slips.}}
	 \label{fig:sample_preparation}
  \end{center}
\end{figure}

The samples used in our experiments were live yeast cells dispersed in a three-dimensional agarose gel. The yeast cells are diluted in a phosphate buffer saline (PBS) solution at the desired concentration at room temperature. A water solution containing 0.7\% agarose is heated up to 80$^\circ$C and then left to cool down to 35$^\circ$C at which point a drop of the PBS solution containing the yeast cells is added. The resulting solution is placed between cover slips and left to cool down to room temperature where it solidifies (becomes like jello). This results in a stable three-dimensional arrangement of cells. Yeast cells remain alive for at least 24 hours in agarose. The distance between the cells can be adjusted by varying the initial concentration of cells. The refractive index of agarose is virtually the same as water (n=1.34 at 406nm). The index of refraction of the yeast cells has an index difference to the agarose on the average $\Delta n$ = 0.07. In Fig.~\ref{fig:sample_preparation}, we show the process for preparing the cell cluster and a typical sample under wide field illumination in a standard microscope.

\begin{figure}[h!]
  \begin{center}
    \includegraphics{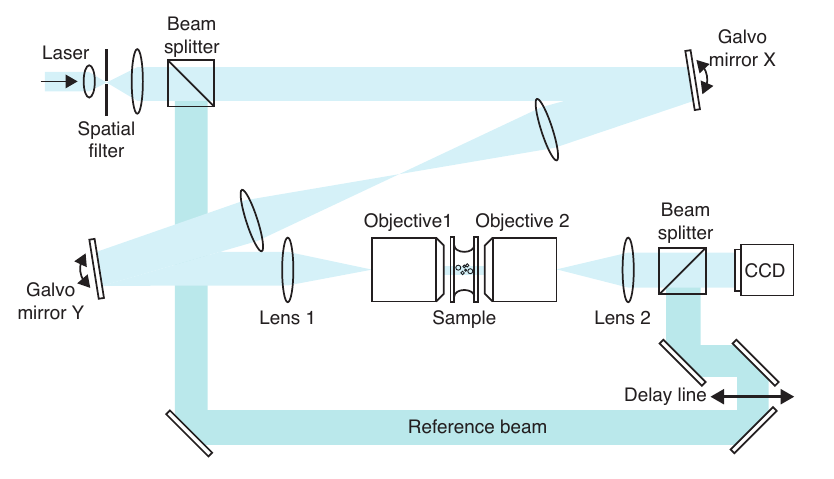}
     \caption{\small{\textbf{Experimental apparatus.} The laser beam is first expanded and passed through a spatial filter in order to produce a clean plane wave. It is then split into a reference and a signal beam. The signal beam is sent on a first galvo-mirror the surface of which is imaged onto a second galvo-mirror through a 4f lens system. The beam is focused in the back focal plane of the illumination objective, which leads to a plane wave illumination of the sample. The sample is imaged on a CCD camera using an objective and a tube lens. The surface of the galvo-mirrors, the sample and the camera plane are in conjugate image plane of each other. The reference beam passes in a delay line to adjust the path length and then recombined in order to create an interference pattern on the camera detector.}}
	 \label{fig:experimental_setup}
  \end{center}
\end{figure}

The experimental apparatus we used for ODT measurements is shown in Fig.~\ref{fig:experimental_setup}. The set-up is a standard optical tomography systems that has been described before~\cite{devaney:2012, kamilov:2015}. The sample is placed between two objectives (Olympus UPlanSApo NA1.40 on the collection side, and Olympus UPlanFI NA1.30 on the detection side) with a working distance in oil of 130$\mu$m and 200$\mu$m respectively, not including the cover slips. The thickness of each of cover slips was 150$\mu$m.  This leaves enough room for a sample with thickness up to 260$\mu$m in water (30 microns are accounted for the oil between the lenses and the cover slips). We tested samples of up to 40$\mu$m in thickness corresponding to 7 or 8 layers of yeast cells.

The light source is a continuous wave diode laser at 406nm, with a coherence length of approximately 250 microns. The beam is spatially filtered, expanded and focused in the back focal plane of the illumination objective thus projecting a plane wave on the sample. The sample is imaged with a lateral magnification of 111 through the collection objective and a relay lens on a detector array (Andor Neo EMCCD, pixel size 6.5 microns), where a hologram is formed through interference with the reference beam. For each view, we measure the total field with the object in place. We then slide the sample together with the cover slips to the side so that the field of view becomes clear of any object. We record then a second field, which give us a measurement of the incident field. This technique allows us to compensate aberrations present in the optical system.

A total of 160 holograms are recorded for each sample. The angle of the reference beam with respect to the optical axis of the signal beam is 35$^\circ$. The reference beam is rotated in a full circle in the transverse plane by the two galvo mirrors in 160 equal increments. The complex field distributions extracted from the holographic recording, are digitized with 16 bits resolution and stored. This data is then ready to be processed to produce the 3D index distribution of the sample.

\begin{figure}[h!]
  \begin{center}
    \includegraphics{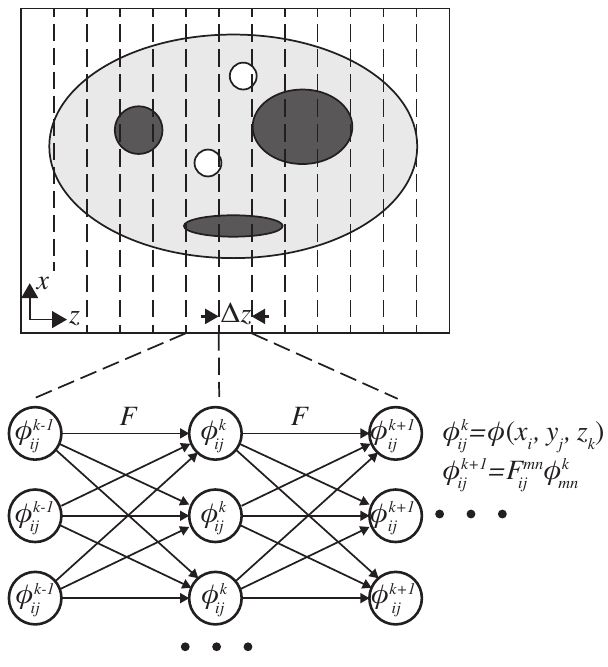}
     \caption{\small{\textbf{Principle of the Beam Propagation Method.} In the propagation process, the space is discretized in $x$, $y$ and $z$. The refractive index inhomogeneity in a slice $z = zk = k\Delta z$ is represented by a phase mask $\phi k(x_i, y_j)$. The BPM can be seen as a neural network where the nodes are the phase retardations $\phi k_{kij}$ and the weights the coefficients of the Fresnel propagation operator, represented by $F$.}}
	 \label{fig:bpm}
  \end{center}
\end{figure}

\section{Results}

\begin{figure}[h!]
  \begin{center}
    \includegraphics{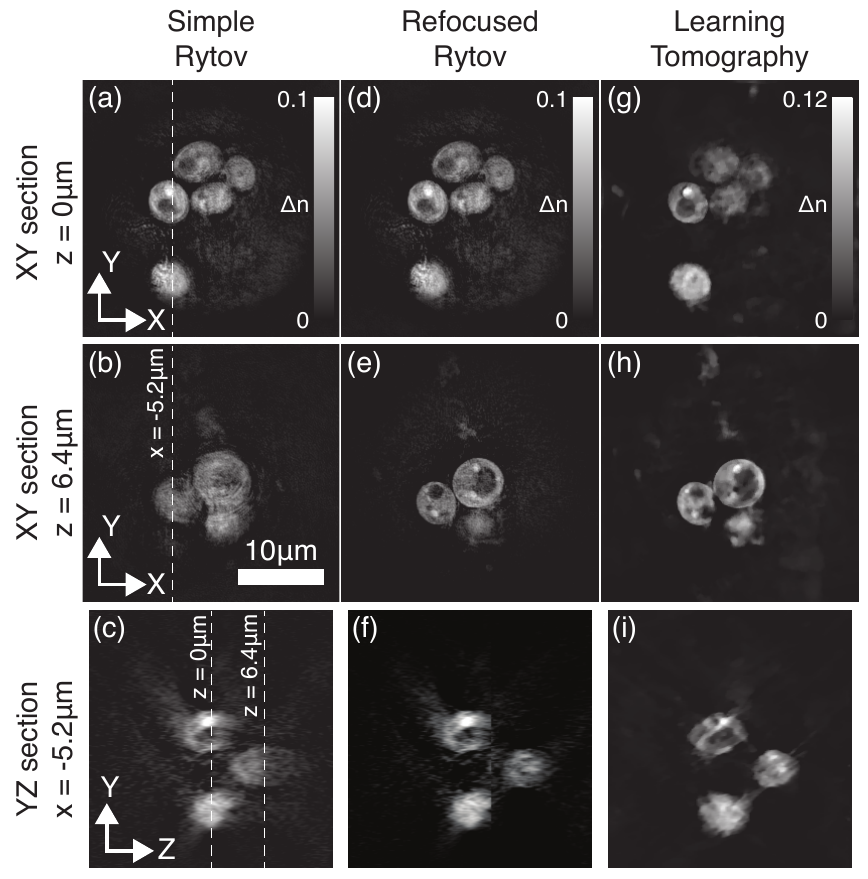}
     \caption{\small{Reconstructions of a cell cluster using the Rytov approximation (a to c), the refocused version of Rytov (d to f), and the Learning Tomography method (g to i). The dashed lines in (a) and (b) indicates the trace of the $y-z$ sections shown in (c), (f) and (g). The two dashed lines in (c) indicate the trace of the $x-y$ planes at two different depth in $z$, i.e. images (a), (d) and (g) at $z = 0\mu$m, and images (b), (e) and (h) at $z = 6.4\mu$m. All the pictures are at the same scale.}}
	 \label{fig:small_group}
  \end{center}
\end{figure}

The images in Fig.~\ref{fig:small_group} display the reconstruction of the 3D index of refraction of a sample consisting of 2 layers of yeast cells with a total thickness less than 20 $\mu$m. Three different reconstruction methods are shown. The first is based on the Rytov approximation. The second column presents the results obtained with the refocused Rytov method. The results presented in Fig.~\ref{fig:small_group} demonstrate that refocused Rytov extends the size of the reconstructed object in the $z$ direction (the depth of field). Finally, the results obtained using Learning Tomography (LT), a nonlinear forward model accounting for multiple scattering, are also shown. The top row in Fig.~\ref{fig:small_group} (Figs. 4a, 4d and 4g) show the reconstruction of the slice of the object that was situated in the focal plane of objective 2 in Fig.~\ref{fig:experimental_setup}. In other words, the image of this plane was in focus on the detector when the holograms were recorded.

\begin{figure}[h!]
  \begin{center}
    \includegraphics{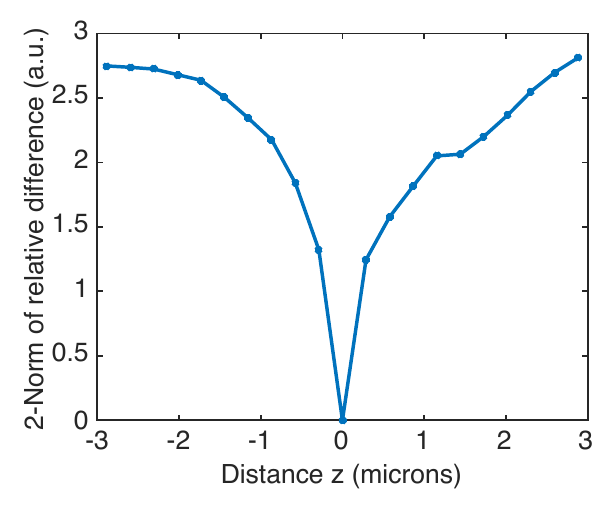}
     \caption{\small{Error in the reconstruction using Rytov approximation as the distance with the imaging plane increases, for the sample shown in Fig.~\ref{fig:bpm}. The quantity plotted is $\|R(z)-RR(z)\|^2/\|RR(z)\|^2$, where $R(z)$ is the Rytov reconstruction (expressed as the refractive index contrast) in slice $z$ (with the imaging plane placed at $z=0$) and $RR(z)$ the Rytov reconstruction refocused in plane $z$ (i.e., with the imaging plane placed at $z$).}}
	 \label{fig:rytov_error}
  \end{center}
\end{figure}

The second row (Figs. 4b, 4e, and 4h) show the reconstruction of a slice of the sample 6.4 $\mu$m away from the plane of best focus. The simple Rytov reconstruction becomes distorted at this position. For ODT algorithms (such as the Born and LT algorithms) which use the detected complex field to form the 3D image of the object, the detected signal can be refocused through post-detection digital propagation. In particular, this is automatically accomplished as part of the LT reconstruction algorithm. The Rytov method on the other hand, requires that the complex phase is extracted from the hologram. For a slice away from $z=0$, the portion of the detected signal that is due to scattering from this slice is defocussed and so is the measured phase. The defocused phase extracted from slices at $z=0$ cannot be digitally refocused as part of the post-detection image formation step since the phase does not obey the wave equation. This accounts for the distortion observed in the Rytov reconstruction (Fig.~\ref{fig:rytov_error}) away from the plane of best focus. 

The defocusing of the Rytov method can be avoided if we physically refocus the optical system multiple times and record the entire sequence of projections for each position. This allows us to reproduce the entire object bystiching together the reconstructions from different locations along $z$ with an accompanying increase in the computational cost and the acquisition time of the data. Fortunately, it is not necessary to physically refocus the system since we have access to the complex field from the holographic recording. The refocusing can be done digitally as a post-detection step followed by a complete Rytov reconstruction for each focal position followed by stitching together of the reconstructions. The images shown in the second row of Fig.~\ref{fig:small_group} were obtained using this refocused Rytov method.  We can obtain an empirical estimate for the depth of focus of the simple Rytov reconstruction by calculating the average squared difference between the simple and refocused Rytov reconstructions. The calculation in Fig.~\ref{fig:rytov_error} can be used to guide the selection of the number of positions in $z$ that need to be selected for refocusing.

The third column in Fig.~\ref{fig:bpm} displays the images obtained with the LT method~\cite{kamilov:2015},~\cite{kamilov:2016}. Unlike the Rytov method, LT does not use the approximation that the detected signal is the result of a single scattering of the illuminating beam. With reference to Fig.~\ref{fig:rytov_error}, each slice acts as a thin transparency modulating the light incident on it. The propagation from one slice to another is accounted for by a free space propagation step. Given the current estimate of the 3D complex index of the object, the BPM gives a prediction for the field incident on the detector. This prediction is then compared to the experimental measurement and an optimization method similar to what is used in multi-layer neural networks gives the estimate for the 3D distribution of the complex index of refraction. The images obtained with LT do not have any defocusing since the propagation model accounts for the entire propagation through the sample. 

The conclusions we draw from Fig.~\ref{fig:small_group} is that for samples consisting of two cells (optical path up to 10 radians and a thickness of 10 microns) refocused Rytov and LT give comparable image quality. For the thickness and complexity (index contrast) of the samples used in the experiments in Fig.~\ref{fig:bpm} the main difference between the results here and the single cell ODT experiments that have been reported extensively in the literature is the depth of focus limitations observed for the Rytov approximation and the solutions provided by refocused Rytov and LT.

\begin{figure}[h!]
  \begin{center}
    \includegraphics{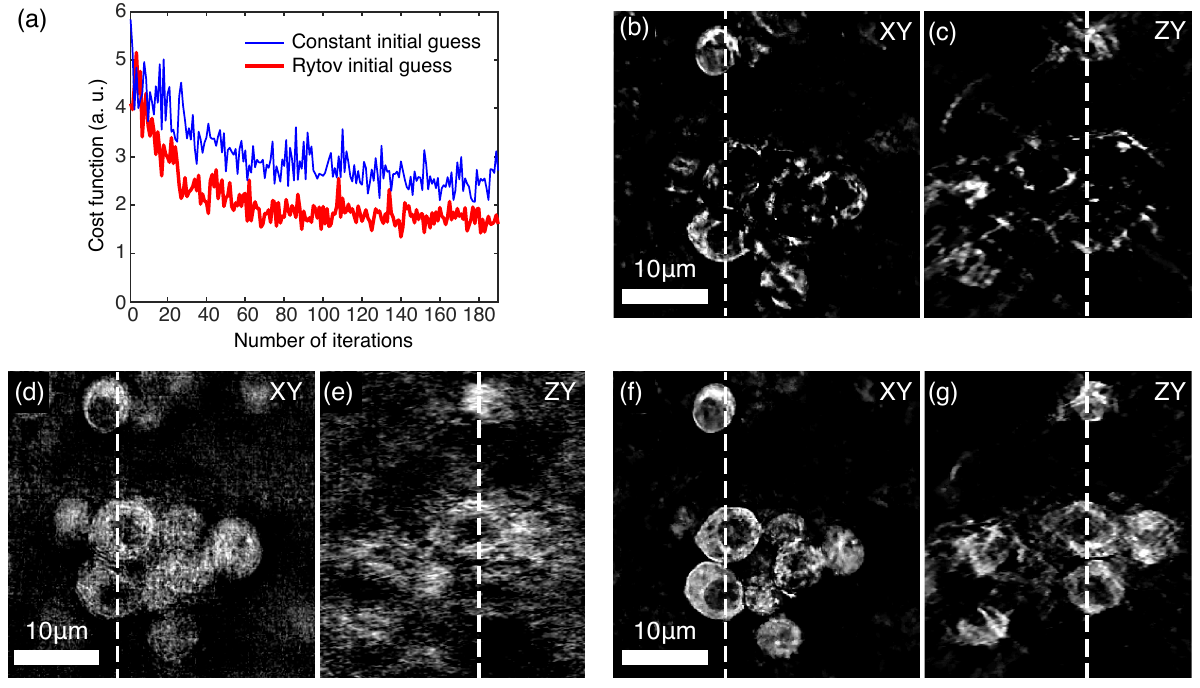}
     \caption{\small{\textbf{Importance of initial guess.} (a) Cost function as a function of the number of iterations for the LT algorithm initialized with a constant refractive index contrast (thin blue curve) of $\Delta n=0.03$, and the simple Rytov reconstruction (thick red curve). (b) XY slice through the LT reconstruction initialized with the constant initial guess. (c) ZY slice of the reconstruction shown in (b). (d) Slice through the Rytov reconstruction. (e) ZY slice of the Rytov. (f) XY slice through the LT reconstruction initialized with the Rytov reconstruction shown in (d) and (e). (g) ZY slice through the LT reconstruction initialized with the Rytov reconstruction. All slices are at $z$ = 3.6 microns from the experimental imaging plane. The gray level is proportional to the refractive index contrast, black corresponding to the background value of $n = 1.34$. In (b, d, f), the vertical dashed line shows the intersection of the ZY section plane of images (c, e, g), respectively. Similarly, in (c, e, g), the vertical dashed line shows the intersection of the XY section plane of images (b, d, f), respectively.}}
	 \label{fig:initial_guess}
  \end{center}
\end{figure}

We have shown in a recent paper~\cite{kamilov:2016} that LT can be more accurate in imaging complex samples consisting of dielectric spheres and cylinders. Here we explore the sample complexity (thickness and index contrast) that LT can reach with objects consisting of cell clusters. The cluster of yeast cells in the experiment of Fig.~\ref{fig:initial_guess} consists of a maximum of 5 layers of cells and has a total thickness of approximately 30 $\mu$m. When LT is initialized with an initial guess of a constant refractive index, the algorithm reaches a local minimum that is highly distorted. When the same cluster is reconstruction with LT but with the simple Rytov initial guess, the result is a more complete and sharper image. Yeast cells are known to have vacuoles (compartments filled with water) and these are clearly visible in Fig.~\ref{fig:initial_guess}(f) but not in Fig.~\ref{fig:initial_guess}(b). Similarly, the y-z image of the LT reconstruction (Fig.~\ref{fig:initial_guess}(g)) displays with high definition the cell structure along the depth of the sample. In general, as the sample complexity increases, the quality of the Rytov reconstruction gradually degrades, whereas LT is less predictable due to the randomness of the local minima in which the algorithm gets stuck.  The better initial condition provided by Rytov gives LT a running start that generally improves the image quality even though this is not guaranteed.

The Rytov approximation becomes distorted for slices in the z-direction more than $\pm 1\mu$m away from the plane of best focus (see Fig.~\ref{fig:rytov_error}). This can be overcome if we use refocused Rytov which compensates for this effect by digitally refocusing the system for multiple distances in $z$ and stitching together the individual solutions to form a complete reconstruction.

\begin{figure}[h!]
  \begin{center}
    \includegraphics{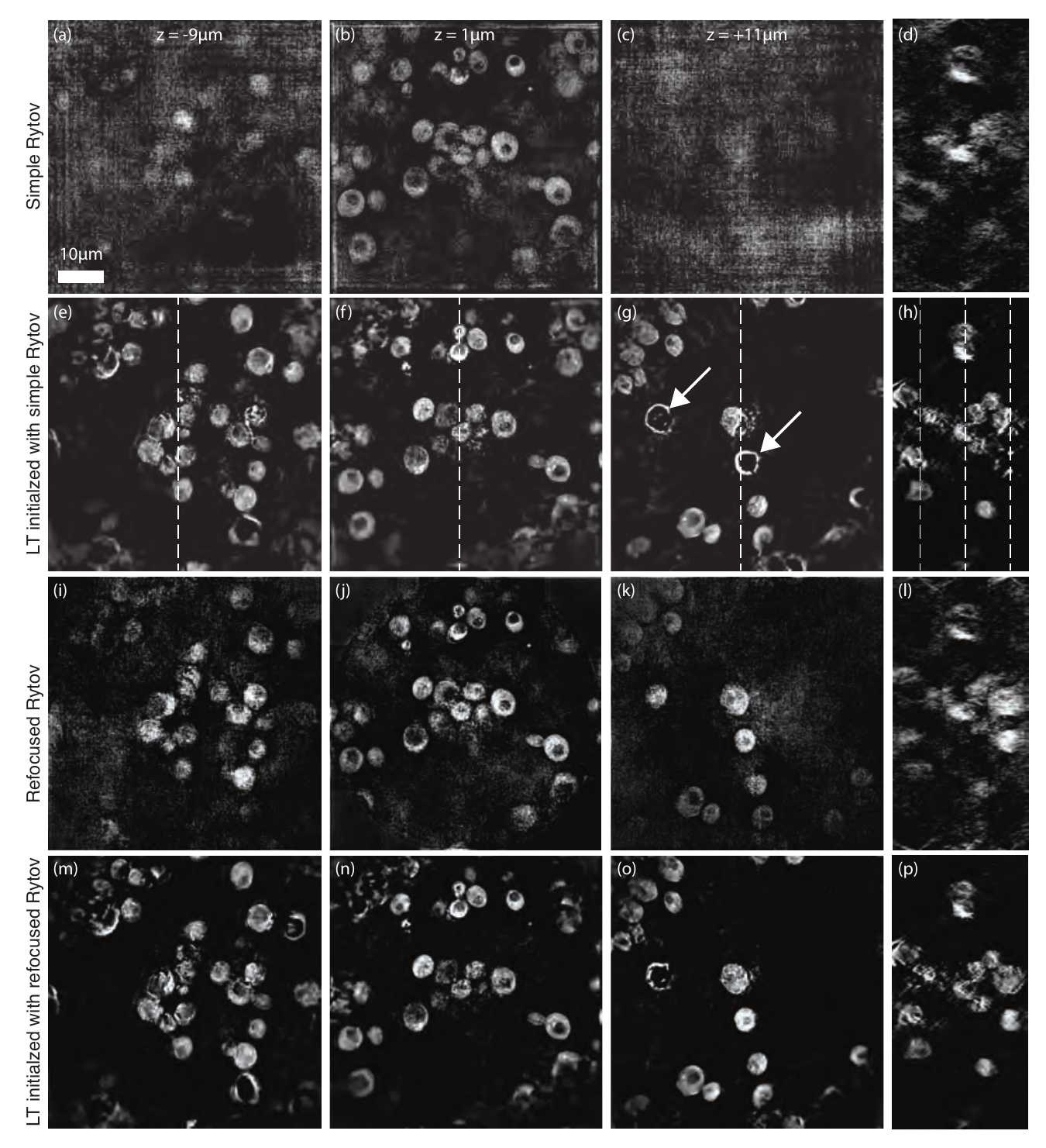}
     \caption{\small{\textbf{Comparison for a 5 layer thick sample of the LT reconstructions starting from two different initial guesses.} (a-c) XY slices through the simple Rytov reconstruction at $z=-9$, 0, +11 microns respectively (imaging plane at $z=0$). (d) ZY section through the simple Rytov reconstruction. (e-g) XY slices through the LT reconstruction initialized with the simple Rytov reconstruction, in the same planes as images a to c. (h) ZY section through the LT reconstruction initialized with the simple Rytov reconstruction. The vertical dashed lines showing the intersection of the XY section with the ZY section. (i-k) XY sections through the stitched multi-domain refocused Rytov reconstruction, refocused in the displayed at $z=-9, 0, +11$ microns. (l) ZY section of the stitched multi-domain refocused Rytov reconstruction. (m-o) XY slices at the corresponding depths of the LT reconstruction initialized with the refocused Rytov reconstruction. (p)  ZY section through the LT reconstruction initialized with the refocused Rytov reconstruction.}}
	 \label{fig:large_group}
  \end{center}
\end{figure}

The results shown in Fig.~\ref{fig:large_group} are images of the same sample as in Fig.~\ref{fig:initial_guess}. The top row is the simple Rytov reconstruction clearly showing the limited depth of field of the technique. The second row shows the LT images obtained when initialized with the Rytov image of the top row and the third row shows the same slices of the cell cluster produced by the refocused Rytov method. In both cases a dramatic improvement in the clarity and contrast of the images is observed. The LT image (second row) has better contrast, sharpness and segmentation of the cells but artifacts due to the local minima are produced in some cases. Finally, the LT image obtained with the refocused Rytov as the initial condition is displayed in the bottom row. The image in the second row is quite similar to the image in the fourth row demonstrating that initialization of the LT algorithm with the simple Rytov is sufficient. Fig.~\ref{fig:large_group}g (the LT reconstruction with simple Rytov initialization) contains two cells (marked with arrows) that appear edge enhanced. Such artifacts are due to local minima in which the LT algorithm was trapped. This is made evident by observing Fig.~\ref{fig:large_group}o (the LT reconstruction with refocused Rytov as the initial condition). In this case the proper contrast of one of the edge enhanced cells was restored because its $z$ position happened to coincide with one of the Rytov refocused planes. Once initialized with the proper shape the algorithm stays there, indicating the presence of a strong minimum.

\section{Conclusion}

We have experimentally explored the capabilities of the refocused Rytov and the Learning Tomography algorithms in thick samples consisting of yeast cell clusters. The LT algorithm yields satisfactory results in term of sharpness and background noise both when starting from the single Rytov. LT has the capability of filling gaps in the initial guess over a range larger than the range of validity of the Rytov approximation (as shown in Fig.~\ref{fig:rytov_error}). Moreover, even in planes in which the Rytov solution has been refocused, generally LT yields sharper images. We believe this is because multiple scattering is accounted for with LT. For objects that are thick, but weakly scattering, the Rytov solution can be quite close to the LT solution. However, for the Rytov reconstruction to be of comparable quality, the measurement needs to be refocused on at least one plane within the range of validity of the approximation, which is around 1 micron in this type of sample. In the case of the sample shown in Fig.~\ref{fig:large_group}, we would thus need 30 refocused Rytov solutions. The phase unwrapping~\cite{bioucas:2007} that is part of each Rytov reconstruction generally dwarfs the computational cost of the optimization in the LT algorithm. Therefore, refocused Rytov is generally much more expensive in terms of computation time.

\bibliographystyle{ieeetr}
\bibliography{thick_samples_tomography}

\end{document}